%% file: main.tex
\documentclass[conference]{IEEEtran}
\IEEEoverridecommandlockouts
\usepackage{cite}
\usepackage{amsmath,amssymb,amsfonts, spconf}
\usepackage{algorithmic}
\usepackage{graphicx}
\usepackage{textcomp}
\usepackage{xcolor}
\def\BibTeX{{\rm B\kern-.05em{\sc i\kern-.025em b}\kern-.08em
    T\kern-.1667em\lower.7ex\hbox{E}\kern-.125emX}}

\usepackage{booktabs}
\usepackage{tabularx} 

\input{includes/macro.tex}
\input{includes/packages.tex}

\begin{document}

\title{CAT: Causal Audio Transformer for Audio Classification
}

\name{Xiaoyu Liu$^1$, Hanlin Lu$^2$, Jianbo Yuan$^2$, Xinyu Li$^2$}
\address{$^1$University of Maryland, College Park, $^2$ByteDance}

\maketitle

\begin{abstract}
The attention-based Transformers have been increasingly applied to audio classification because of their global receptive field and ability to handle long-term dependency. However, the existing frameworks which are mainly extended from the Vision Transformers are not perfectly compatible with audio signals. In this paper, we introduce a Causal Audio Transformer (CAT) consisting of a Multi-Resolution Multi-Feature (MRMF) feature extraction with an acoustic attention block for more optimized audio modeling. In addition, we propose a causal module that alleviates over-fitting, helps with knowledge transfer and improves interpretability. CAT obtains higher or comparable state-of-the-art classification performance on ESC50, AudioSet and UrbanSound8K datasets, and can be easily generalized to other Transformer-based models. 
\end{abstract}


\input{sections/0_introduction.tex}

\input{sections/1_related_work.tex}

\input{sections/2_1_model_arch.tex}

\input{sections/2_method.tex}

\input{sections/3_experiments.tex}

\input{sections/5_conclusion.tex}

\bibliographystyle{IEEEbib}
{\small \bibliography{reference}}

\end{document}

%% file: includes/macro.tex
\newcommand{\Transformer}{Transformer\xspace}
\newcommand{\Transformers}{Transformers\xspace}

\newcommand{\ourmodel}{Causal Audio Transformer\xspace}
\newcommand{\modelshortcut}{CAT\xspace}

\DeclareMathOperator*{\argmin}{\arg\!\min}

%% file: includes/packages.tex
\usepackage{xspace}
\usepackage{tablefootnote}
\usepackage{subcaption}
\usepackage{makecell}
\usepackage{wrapfig}
\usepackage{graphicx}
\usepackage{booktabs}
\usepackage{xcolor}
\usepackage[colorlinks = true,
            linkcolor = blue,
            urlcolor  = blue,
            citecolor = blue,
            anchorcolor = blue]{hyperref}
\usepackage{cleveref}

%% file: sections/0_introduction.tex
\section{Introduction}
Audio classification is the task of analyzing audio recordings and assigning the corresponding labels to the audio sample. Having witnessed the huge success of \Transformers\cite{vaswani2017attention} in the field of Natural Language Processing (NLP) and Computer Vision (CV), they have been adapted to the audio modality and have obtained the state-of-the-art performances. Part of the success is attributed to the global receptive fields in \Transformers to capture long-range context in the audio signals. 
The existing audio transformer models inherit the structure of the famous Vision \Transformer (ViT) \cite{dosovitskiy2020image}, mainly because one of its most widely-used features, the Mel-Spectrogram, is of the same format as an image. Instead of letting the x- and y-axis carry the spatial information in image modeling, the x-axis of the Mel-Spectrogram denotes temporal information while the y-axis carries the discrete frequency information for audio inputs\cite{nagrani2021attention, koutini2021efficient, gong2021ast, chen2022hts, gong2021psla, kong2020panns}.

While achieving superior performance, there are still open challenges in these trendy audio \Transformers:
(1) The commonly used acoustic representations leverage different time-frequency transformations and contain acoustic semantics of various scales and different granularities~\cite{chen2022hts,gong2021ast,koutini2021efficient}, which can hardly be effectively captured by ViTs~\cite{dosovitskiy2020image} using conventional self-attentions, patch sampling and embeddings; 
(2) The successful designs in visual tasks, such as resnet~\cite{krizhevsky2009learning}, mixup~\cite{zhang2017mixup}, are shown to be more prone to over-fitting and less generalizable in the acoustic domain; 
(3) Feature selection and representation learning are critical in computer vision, yet have been often overlooked in acoustic modeling.

In the face of the challenges, we propose a \textbf{C}ausal \textbf{A}udio
\textbf{T}ransformer (\modelshortcut), which includes Multi-Resolution Multi-Filter (MRMF) feature extraction, an acoustic attention and a causal module. 
Spectrograms, as standard inputs in audio models, are sampled using Fourier Transform, leading to a natural trade-off between the temporal resolution and the frequency resolution. CAT balances the trade-off by extracting comprehensive temporal-frequency feature patches in multiple resolutions and filters which is later combined with 3D positional embeddings. 
Then, the acoustic attention is proposed to effectively extract semantics from such representations, taking the feature patches as inputs. The patches from different filters are evenly distributed among the attention heads, while we calculate the pair-wise attentions among patches from different resolutions but within the same time frames, allowing information exchange in various granularities.
We further introduce a causal module to establish the necessary and sufficient relationship between the learned representation and predicted labels based on counterfactual reasoning\cite{pearl2009causality, wang2021desiderata}. We extend \cite{wang2021desiderata} to the context of audio classification where a lower bound of Probability of Necessity and Sufficiency (PNS) in terms of interventional distribution is provided. Since such a lower bound can only be estimated from the true distribution, we propose a causal module that learns a mapping from the interventional distribution to the observational dataset (i.e., the one we have), to alleviate over-fitting, improve interpretability and enable better knowledge transfer.

CAT achieves SOTA performance on ESC50~\cite{piczak2015esc}, AudioSet~\cite{gemmeke2017audio} and UrbenSound8K~\cite{salamon2014dataset}.
To sum up, our contributions are: 
\begin{itemize}
    \item \ourmodel (CAT), with \textit{Multi-Resolution Multi-Filter (MRMF)} features extraction and  \textit{an acoustic attention} for acoustic modeling. 
    \item Causal loss with reconstruction block, that explicitly measures the feature quality using Probability of Necessity and Sufficiency (PNS), alleviates over-fitting, and improves knowledge transfer across different datasets.
\end{itemize}

\input{tex_img_tab/training.tex}

%% file: tex_img_tab/training.tex
\begin{figure*}[!ht]
    \centering
    \includegraphics[width=0.95\textwidth]{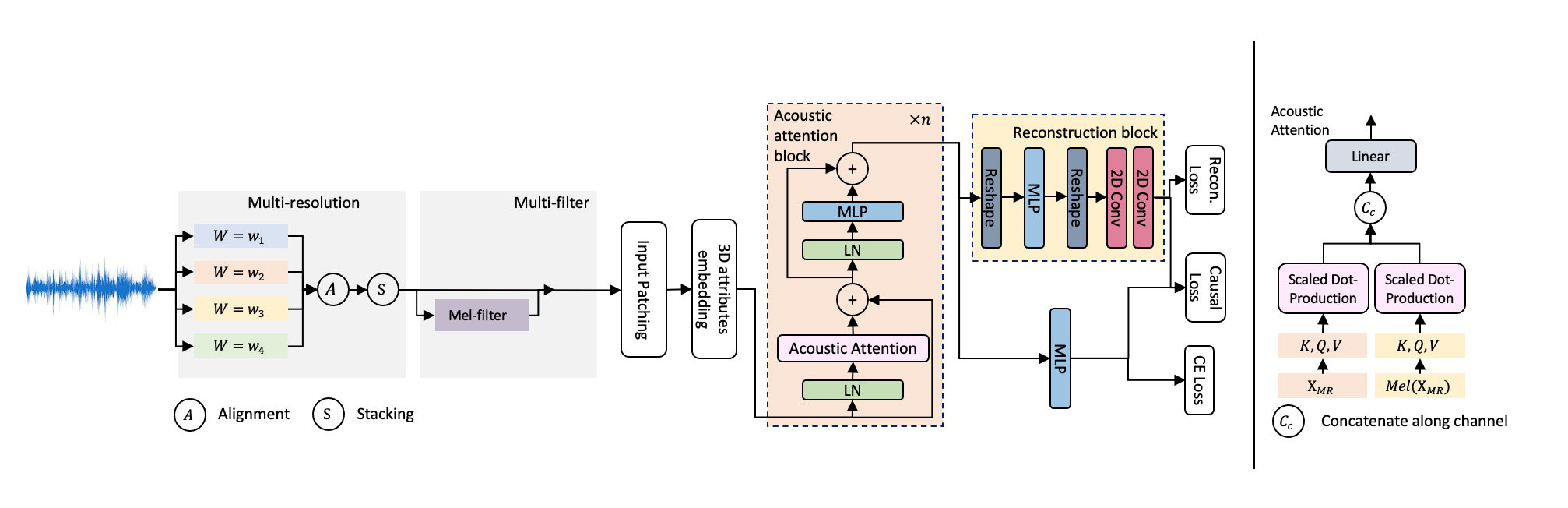}
     \vspace{-11mm}
    \caption{Left: Overview of proposed CAT; Right: the detailed structure of proposed acoustic attention. }
    \label{fig:training}
    \vspace{-3mm}

\end{figure*}

%% file: sections/1_related_work.tex
\section{Related Work}
\noindent{\textbf{Audio classification: from CNN to Transformer.}}
Audio classification is the task of predicting labels of soundtracks. The great success of large-scale training originates from the rise of CNN models in audio classification\cite{kong2020panns, lopez2021efficient, kim2020urban, nanni2021ensemble, sailor2017unsupervised, huang2018aclnet}. More recently, with attention mechanism triumphs in the field of Natural Language Processing (NLP) and Computer Vision (CV), self-attention-based transformers are first applied to acoustic classification by AST~\cite{gong2021ast}, where the spectrogram is used as the input of a ViT~\cite{dosovitskiy2020image} backbone. Multiple subsequent studies focus on improving the model efficiency: PaSST~\cite{koutini2021efficient} proposes a patch-out mechanism, and HTS-AT\cite{chen2022hts} adopts a hierarchical backbone~\cite{liu2021swin}. To further boost the performance, MBT\cite{nagrani2021attention} introduces an visual-acoustic fusion and PLSA\cite{gong2021psla} introduces a model-agnostic framework. However, the network structures in most recent studies are heavily borrowed from vanilla transformers originally proposed for NLP and CV tasks, and are more prone to over-fitting and less generalizable to acoustic data. Therefore, in this paper, we introduce the acoustic attention which incorporates MRMF features specialized for acoustic modeling. The proposed acoustic attention can be easily extended to various \Transformer blocks.

\noindent{\textbf{Causal inference in feature representation.}}
The concept of causality is first introduced in graphical probabilistic models\cite{pearl2009causality}. Although causal inference is a relatively new concept in audio classification, it has shown advances in interpretable machine learning and representation learning. Modeling causality among generating factors significantly encourages the learning of representative features\cite{yang2021causalvae}. \cite{nangi2021counterfactuals} adopts counterfactual information that helps knowledge transfer across different domains. \cite{wang2021desiderata} proves a lower bound of the learned representation being a necessary and sufficient condition for label prediction. However, such a lower bound is in the interventional setting, and thus cannot be estimated directly without knowing the true distribution. We base our design on \cite{wang2021desiderata} and extend it to a causal module that maps the lower bound to the observational dataset in the context of audio classification.

%% file: sections/2_1_model_arch.tex
\section{Causal Audio Transformer}
In this section, we introduce our Causal Audio Transformer (CAT). CAT first extracts MRMF feature patches with 3D positional embedding. The feature patches are then sent as input to the acoustic attention (Section~\ref{sec:model-multi}). Then a causal module is proposed to alleviate over-fitting and improve interpretability and knowledge transfer (Section~\ref{sec:causal}).

\subsection{Transformer For Acoustic Modeling}
\label{sec:model-multi}
Image modeling is largely different from acoustic modeling because a 2D image only carries spatial information along its axis, whereas acoustic features convey both temporal and frequency information. Therefore, it is not trivial to apply the conventional vision transformer for efficient acoustic modeling. We bridge the gap by extracting MRMF features and acoustic attention. 

\noindent\textbf{Multi-resolution multi-filter feature extraction.} 
Due to the nature of 1D Fourier Transformation (FT), there is a trade-off between frequency and temporal resolution in a temporal-frequency representation (e.g. spectrogram). We propose to extract spectrograms of different temporal resolutions as:
\begin{equation}
    x_{\text{MR}} = \{\text{FFT}_{\omega}(s)\}, \omega\in \{\omega_1,\omega_2,...,\omega_K\}
\end{equation}
where s denotes the input audio wave and $\text{FFT}_{\omega}$ denotes the 1D FFT with a window size of $\omega$, $\{*\}$ represents a set.  For each raw input $x$, $x_{\text{MR}}$ contains k spectrograms, each of size $\mathrm{R}^{T_i\times F}, i = 1, 2, ..., K$.
$T_i$ represents the number of time frames, and $F$ is the number of frequency bins. 
Such multi-resolution spectrograms share similar intuitions with~\cite{zhu2018learning, grais2018multi}, but ours is paired with acoustic attention and 3D positional embedding, as introduced in the next section, so that the information can be more effectively taken use of.
We further allow  spectrograms under multiple filters to serve as parallel inputs to different attention heads. For example, we could pass a copy of $x_{\text{MR}}$ to the Mel-filter bank, obtaining the MRMF features $x_{\text{MRMF}} \in \mathbb{R}^{\{T_i\}\times K\times F \times 2}$ as follows:
\begin{equation}
    x_{\text{MRMF}}= \{\text{mel}(x_{\text{MR}}),x_{\text{MR}}\}
\end{equation}

Spectrograms of different resolutions have different sizes. There are multiple ways to align them. One possibility is to concatenate them along the temporal dimension, forming a super spectrogram. Another method is to project the spectrograms to the same temporal dimension. We adopt the later setting based on better empirical results. \\

\noindent\textbf{Patching and 3D positional embedding.} 
The input $x_{\text{MRMF}}$ are patchified and aggregated with windows ${Wi} \in \mathrm{R}^{1\times K\times F\times 1}$, and we use linear projections $\{\xi\}$ to match the dimension to $M$. 
Different from the square-shaped input patching in computer vision \cite{dosovitskiy2020image}, the proposed patch $x_p$ contains frequency information of various resolutions. Such design aggregates and preserves the important multi-scale frequency information, which is critical for audio classification. 
Inspired by 3D positional embedding used in video transformer~\cite{zhang2021vidtr}, we propose an acoustic 3D positional embedding to make the network aware of features extracted with different window sizes and filters: 
\begin{equation}
    pe = (g([pe_1, pe_3])\otimes \mathbf{1}_F )^T + pe_2\otimes \mathbf{1}_T
\end{equation}
where $g$ is a linear projection, $pe_1$ is a $T$ dimensional sinusoid time embedding and $pe_2$ a $F$ dimensional frequency sinusoid embedding, following \cite{vaswani2017attention}. $pe_3$ is the one-hot encoding of resolutions. $\mathbf{1}_F$ is $F$-dimensional all-one vector.\\

\noindent\textbf{Acoustic attention.} We consequently design the acoustic attention to incorporate MRMF feature patches in audio \Transformers. Specifically, multi-filtered inputs are fed into different heads and processed with separate scaled dot-production (Figure \ref{fig:training}) so that attentions with and without Mel-filtering are calculated independently. Note that the proposed acoustic attention works with various attention kernels (e.g. multi-head attention~\cite{vaswani2017attention}, SWIN-attention~\cite{liu2021swin}, etc.). We use the SWIN kernel for our experiments.

%% file: sections/2_method.tex
\subsection{Causal module for better representation}
\label{sec:causal}
The idea of encouraging the learned hidden representation to be a necessary and sufficient cause of label prediction is first proposed in \cite{wang2021desiderata}, by maximizing a lower bound under intervention. \cite{wang2021desiderata} demonstrated it helps build models with less over-fitting and more robustness. However, such a bound requires an estimation of the unobserved confounder $C$, making it hard to succeed in empirical setups. In this paper, we introduce a causal module with a causal loss that applies addition constraints on bijective mappings to resolve the issue.  In this section, we first introduce the preliminaries from a causal perspective, then introduce the proposed module.

\noindent{\textbf{Preliminaries}} Let $X$ denote the training data and $Y$ be the counterfactual label function. we define a function $f$ that maps $X$ to its latent representation $Z$, where $Z = f(X)$. 
The probability of sufficiency and necessity of $\mathbb{I}\{Z = z\}$ for $\mathbb{I}\{Y = y\}$ is defined as \cite{wang2021desiderata}:
\begin{equation}
    \text{PNS}_{Z=z, Y=y} = P(Y(Z \neq z) \neq y, Y(Z = z) = y).
\end{equation}
where $Y(Z=z)$ reasons about what the label would be if $Z=z$. 
To combat over-fitting and ensure the feature is representative, our goal is to maximize the PNS during training. \cite{wang2021desiderata} established the  lower-bounded of PNS as:
\begin{equation}
\label{eq:bound}
    \text{PNS} \geq P(Y=y | do(Z=z)) - P(Y\neq y | do(Z \neq z))
\end{equation}
where $do$ notation\cite{pearl2009causality} stands for intervention of $Z$.

However, without knowing the true distribution, the corresponding label change under the certain intervention of latent representation $Z$ cannot be directly estimated. Instead of solving the problem through deconfounding, we derive an estimation of the lower bound as:
\begin{equation}
    \label{eq:final}
    \footnotesize
    \text{PNS} \geq = \int P(Y|X) [P(X|f(X)=z) - P(X|f(X)\neq z)] dX
\end{equation}
We could observe from a probabilistic view that maximizing $P(Y|X)$ is the classification objective, and $P(X|f(X)=z)$ is the probability of inferring $X$ from a specific latent representation $Z=z$. Given that we want to maximize such a lower bound of PNS, a feasible strategy is to maximize both $P(Y|X)$ and $P(X|f(X)=z) - P(X|f(X)\neq z)$. The former is consistent with the classification objective, while the second term indicates that we need an accurate and bijective matching from $Z$ to $X$. 

\noindent{\textbf{Reconstruction and causal loss}}
We argue that such mapping can be accomplished by conducting a reconstruction module and a ``causal'' loss, as shown in Figure~\ref{fig:training}.
We purpose a simple reconstruction block that runs in parallel with a classification layer so that $X$ and $Z$ approximate bijective relationship measured by the reconstruction loss $l_{rs}$:
\begin{equation}
     l_{rs} = \lVert \Phi(Z) - X \rVert_2 
\end{equation}
where $\Phi$ is parameterized by a reconstruction block as in Figure~\ref{fig:training}. 

With the underlying requirement being satisfied, we further minimize the causal loss ( \cite{wang2021desiderata} Equation 25 )as :
\begin{equation}
    lc = -\sum_{j=1}^d log \text{PNS}_n (f_j(X), Y | f_{-j}(X))
\end{equation}

where 
\begin{equation}
\text{PNS}(f_j(X),Y| f_{-j}(X)) = \prod_{i=1}^n \text{PNS}_{f_j(X_i), Y_i|f_{-j}(X_i)}
\end{equation}

The objective of CAT is the sum of a cross-entropy loss $l_\theta$, a reconstruction loss, and a causal loss:
\begin{equation}
   L= \argmin (l_\theta + l_c + l_{rs}) 
\end{equation}

%% file: sections/3_experiments.tex
\section{Experiments}

\subsection{Dataset}
We evaluate \modelshortcut on three datasets.
\textbf{AudioSet\cite{gemmeke2017audio}}: contains over two million 10-second audio clip with 527 labels. We report mean average precision (mAP) on evaluation set following \cite{gong2021ast}. 
\textbf{ESC50\cite{piczak2015esc}: } contains 50 categories, 2000 audio recordings. We report the averaged accuracy of 5-fold cross-validate and mAP of 5 rounds of experiments following~\cite{gong2021ast}.
\textbf{UrbanSound8K\cite{salamon2014dataset}:}  contains 8732 labeled sound excerpts of 10 classes. We report the average accuracy of 10-fold cross-validation following \cite{gazneli2022end}.

\subsection{Implementation details}

Following previous recipe \cite{kong2020panns,gong2021ast, koutini2021efficient}, we use 32kHz as sample rate for all experiments. We generate both the spectrogram  and mel-spectrograms in a window of 25ms and hop length of 10ms.  In training, we adopt the inverse proportional weighting\cite{izmailov2018averaging} to balance the AudioSet training data, the mixup of 0.5 \cite{tokozume2017learning} is used in all of our experiments.

\subsection{Main Results}

\input{tex_img_tab/tab_audio.tex}
\noindent\textbf{Audioset}
We show CAT performance on AudioSet in Table~\ref{tab:main_result}. Our model significantly outperforms previous work based on ConvNets \cite{kong2020panns}, as well as SOTA transformer-based models by a noticeable margin (+$0.7\%$ comparing to \cite{chen2022hts}). Note that both our CAT and HTS-AT share a similar SWIN backbone, the superior performance demonstrates that our proposed audio-transformer with MRMF and acoustic attention, plus the causal module are effective.  It is worth mentioning that following the same ensemble protocol, the CAT achieves performance comparable to SOTA ensemble models. By simply ensemble the CAT with PANN~\cite{kong2020panns}, we achieve performance comparable to the previous 9-model ensemble~\cite{koutini2021efficient} results. This also shows the CAT provides complementary information to ConvNet based model.  

\noindent\textbf{ESC50}
The CAT achieves SOTA performance on ESC50 dataset as well (Table~\ref{tab:main_result}). We observe a similar pattern that the proposed CAT outperforms previous SOTA \cite{gong2021ast, chen2022hts} by 0.8\% and 0.2\%  with and without ensemble; demonstrating the effectiveness of the proposed design. Note that ensemble CAT with other acoustic model\cite{kong2020panns} achieves performance comparable to previous work trained on additional modality, e.g. CAT is 0.7\% better than AnT~\cite{elizalde2022clap} which utilizes visual-acoustic information. It can be anticipated that incorporating multimodality features into CAT may further improve performance.

\subsection{Ablations}
\input{tex_img_tab/tab_ablation.tex}
We perform ablations on ESC50 and AudioSet.

\noindent\textbf{Effectiveness of causal loss. } Table \ref{tab:abl_causal_loss} show that the causal loss helps with performance on datasets of different scale. \\

\noindent\textbf{Impact of pre-training.}
The imagenet pre-training helps with the performance on both datasets (Table \ref{tab:abl_pre_train}), however, the impact is more significant on ESC50. This is due to that the transformer generally requires large-scale data to train. \\

\noindent\textbf{Generalization}
We show that the proposed MRMF and causal loss generalize well to different backbone (e.g. backbone w \cite{liu2021swin} and w/o \cite{gong2021ast} hierarchy), and consistently improve the performance in Table \ref{tab:generalization}.\\

\noindent\textbf{Acoustic embedding quality}
We show that the proposed causal loss is able to eliminate noise and thus the feature transfer better across datasets (Table \ref{tab:feature}). We perform pre-training on small-scale datasets to ensure the performance gain is from proposed causal loss instead of data overlapping.

%% file: tex_img_tab/tab_audio.tex

\begin{table}[!htbp]
\footnotesize
  \caption{\textit{Evaluation on AudioSet and ESC50 dataset. We report mAP for AudioSet and top1 average accuracy for ESC50.* denotes the ensemble version of the model. }}
  \vspace{-.6cm}
  \label{tab:main_result}
  \resizebox{\columnwidth}{!}{%
  \centering
  \begin{tabular}{llcc}
    \toprule
    \textbf{Models}                         & \textbf{Modality}  & \textbf{AudioSet \%wo} & \textbf{ESC50 \%} \\
    \midrule
    PANN \cite{kong2020panns}       & Audio                &  0.434            & 94.7  \\     
    AST \cite{gong2021ast}          & Audio                &  0.347            & 95.6      \\
    HTS-AT \cite{chen2022hts}       & Audio                &  0.471            & -  \\
    \textbf{CAT}                    & Audio                &  \textbf{0.478}   & \textbf{96.4}     \\

    \midrule
    PSLA* \cite{gong2021psla}              & Audio        &  0.474      & -      \\
    AST*\cite{gong2021ast}                  & Audio       &  0.475      & 95.7      \\
    HTS-AT*\cite{chen2022hts}               & Audio       &  0.487      & 97.0      \\
    PaSST*\cite{koutini2021efficient}       & Audio       &  0.496      & 96.8      \\
    CLAP \cite{elizalde2022clap}        & Audio+Text     &  -        & 96.7      \\
    AnT\cite{zhao2021connecting}         & Audio+Video   &  -        & 95.7      \\
    \textbf{CAT*}                           & Audio       &  \textbf{0.489}   & \textbf{96.9}    \\
    \textbf{CAT + PANN*}                      & Audio       &  \textbf{0.491}   & \textbf{97.2}     \\

  \bottomrule
\end{tabular}
}
\vspace{-4mm}
\end{table}

%% file: tex_img_tab/tab_ablation.tex
\begin{table}[t]
    \footnotesize
        \caption{\textit{Ablation on ESC50 and AudioSet. CL denotes causal loss and US8K denotes urbanSound8K.}}
         \vspace{-8mm}
	\centering
	\subfloat[Causal loss helps with performance on both dataset.]
	{
	\scalebox{0.95}{
		\begin{tabularx}{0.2\textwidth}{lcc}
		\toprule
	    &  {\textit{AudioSet}}  &{\textit{ESC50}} \\
            \midrule
               w/o CL        & 0.474  & 95.9 \\
               w. CL         & 0.478  & 96.4 \\
		\bottomrule	
		\end{tabularx}
		}
		\label{tab:abl_causal_loss}
	}\hfill
	\subfloat[Pre-training helps with performance on both datasets.]{
	\scalebox{0.95}{
		\begin{tabularx}{0.23\textwidth}{lrr}
		\toprule
		Pre-train &  {\textit{AudioSet}}  &{\textit{ESC50}} \\
    			\midrule
               N/A          & 0.451        & 89.4\\
               ImageNet     & 0.478       & 96.4\\
    		\bottomrule	
    		\end{tabularx}
		}
		\label{tab:abl_pre_train}        
        
    }\hfill
    \vspace{2mm}
	\subfloat[The MRMF and CL generalize to different backbone.]{
    \scalebox{0.9}{
		\begin{tabularx}{0.2\textwidth}{lrr}
		\toprule
		& {\textit{AST\cite{gong2021ast}}} & {\textit{CAT}} \\
		\midrule
           -                & 88.6      & 88.7\\
           + MRMF           & 88.9      & 89.1  \\
           + CL             & 89.1      & 89.4\\
		\bottomrule	
		\end{tabularx}
        }
        \label{tab:generalization}
	}\hfill
	\subfloat[The proposed causal loss helps with feature domain adaptation.]{
	\scalebox{0.9}{
		\begin{tabularx}{0.27\textwidth}{lccc}
		\toprule
		{\textit{pre-train}}  &{\textit{val}} & {\textit{w/o CL}} &{\textit{w. CL}} \\
            \midrule
            US8K      & ESC50 & 84.1  & 86.7\\
            ESC50     & US8K  & 95.9  & 96.1\\
            \bottomrule	
            \end{tabularx}
		}
		\label{tab:feature}
	}
 \vspace{-3mm}
\end{table}

%% file: sections/5_conclusion.tex
\section{Conclusion and Future Work}
In this paper, we propose \ourmodel, an acoustic transformer designed for audio classification. CAT has  MRMF features, an acoustic attention and a causal module. We empirically show that the proposed CAT achieves SOTA performance on multiple datasets. Our ablation also demonstrates the MRMF and causal loss generalize well to different backbones.
\newpage

%% file: main.bbl
\begin{thebibliography}{10}

\bibitem{vaswani2017attention}
Ashish Vaswani and et~al.,
\newblock ``Attention is all you need,''
\newblock {\em Advances in neural information processing systems}, vol. 30,
  2017.

\bibitem{dosovitskiy2020image}
Alexey Dosovitskiy, Lucas Beyer, Alexander Kolesnikov, Dirk Weissenborn,
  Xiaohua Zhai, Thomas Unterthiner, Mostafa Dehghani, Matthias Minderer, Georg
  Heigold, Sylvain Gelly, et~al.,
\newblock ``An image is worth 16x16 words: Transformers for image recognition
  at scale,''
\newblock {\em arXiv preprint arXiv:2010.11929}, 2020.

\bibitem{nagrani2021attention}
Arsha Nagrani, Shan Yang, Anurag Arnab, Aren Jansen, Cordelia Schmid, and Chen
  Sun,
\newblock ``Attention bottlenecks for multimodal fusion,''
\newblock {\em Advances in Neural Information Processing Systems}, vol. 34, pp.
  14200--14213, 2021.

\bibitem{koutini2021efficient}
Khaled Koutini, Jan Schl{\"u}ter, Hamid Eghbal-zadeh, and Gerhard Widmer,
\newblock ``Efficient training of audio transformers with patchout,''
\newblock {\em arXiv preprint arXiv:2110.05069}, 2021.

\bibitem{gong2021ast}
Yuan Gong, Yu-An Chung, and James Glass,
\newblock ``Ast: Audio spectrogram transformer,''
\newblock {\em arXiv preprint arXiv:2104.01778}, 2021.

\bibitem{chen2022hts}
Ke~Chen and et~al.,
\newblock ``Hts-at: A hierarchical token-semantic audio transformer for sound
  classification and detection,''
\newblock in {\em ICASSP 2022-2022 IEEE International Conference on Acoustics,
  Speech and Signal Processing (ICASSP)}. IEEE, 2022, pp. 646--650.

\bibitem{gong2021psla}
Yuan Gong, Yu-An Chung, and James Glass,
\newblock ``Psla: Improving audio tagging with pretraining, sampling, labeling,
  and aggregation,''
\newblock {\em IEEE/ACM Transactions on Audio, Speech, and Language
  Processing}, vol. 29, pp. 3292--3306, 2021.

\bibitem{kong2020panns}
Qiuqiang Kong, Yin Cao, Turab Iqbal, Yuxuan Wang, Wenwu Wang, and Mark~D
  Plumbley,
\newblock ``Panns: Large-scale pretrained audio neural networks for audio
  pattern recognition,''
\newblock {\em IEEE/ACM Transactions on Audio, Speech, and Language
  Processing}, vol. 28, pp. 2880--2894, 2020.

\bibitem{krizhevsky2009learning}
Alex Krizhevsky, Geoffrey Hinton, et~al.,
\newblock ``Learning multiple layers of features from tiny images,''
\newblock 2009.

\bibitem{zhang2017mixup}
Hongyi Zhang, Moustapha Cisse, Yann~N Dauphin, and David Lopez-Paz,
\newblock ``mixup: Beyond empirical risk minimization,''
\newblock {\em arXiv preprint arXiv:1710.09412}, 2017.

\bibitem{pearl2009causality}
Judea Pearl,
\newblock {\em Causality},
\newblock Cambridge university press, 2009.

\bibitem{wang2021desiderata}
Yixin Wang and Michael~I Jordan,
\newblock ``Desiderata for representation learning: A causal perspective,''
\newblock {\em arXiv preprint arXiv:2109.03795}, 2021.

\bibitem{piczak2015esc}
Karol~J Piczak,
\newblock ``Esc: Dataset for environmental sound classification,''
\newblock in {\em Proceedings of the 23rd ACM international conference on
  Multimedia}, 2015, pp. 1015--1018.

\bibitem{gemmeke2017audio}
Jort~F Gemmeke, Daniel~PW Ellis, Dylan Freedman, Aren Jansen, Wade Lawrence,
  R~Channing Moore, Manoj Plakal, and Marvin Ritter,
\newblock ``Audio set: An ontology and human-labeled dataset for audio
  events,''
\newblock in {\em 2017 IEEE international conference on acoustics, speech and
  signal processing (ICASSP)}. IEEE, 2017, pp. 776--780.

\bibitem{salamon2014dataset}
Justin Salamon, Christopher Jacoby, and Juan~Pablo Bello,
\newblock ``A dataset and taxonomy for urban sound research,''
\newblock in {\em Proceedings of the 22nd ACM international conference on
  Multimedia}, 2014, pp. 1041--1044.

\bibitem{lopez2021efficient}
Paulo Lopez-Meyer, Juan~A del Hoyo~Ontiveros, Hong Lu, and Georg Stemmer,
\newblock ``Efficient end-to-end audio embeddings generation for audio
  classification on target applications,''
\newblock in {\em ICASSP 2021-2021 IEEE International Conference on Acoustics,
  Speech and Signal Processing (ICASSP)}. IEEE, 2021, pp. 601--605.

\bibitem{kim2020urban}
Jaehun Kim,
\newblock ``Urban sound tagging using multi-channel audio feature with
  convolutional neural networks,''
\newblock {\em Proceedings of the Detection and Classification of Acoustic
  Scenes and Events}, 2020.

\bibitem{nanni2021ensemble}
Loris Nanni, Gianluca Maguolo, Sheryl Brahnam, and Michelangelo Paci,
\newblock ``An ensemble of convolutional neural networks for audio
  classification,''
\newblock {\em Applied Sciences}, vol. 11, no. 13, pp. 5796, 2021.

\bibitem{sailor2017unsupervised}
Hardik~B Sailor, Dharmesh~M Agrawal, and Hemant~A Patil,
\newblock ``Unsupervised filterbank learning using convolutional restricted
  boltzmann machine for environmental sound classification.,''
\newblock in {\em InterSpeech}, 2017, vol.~8, p.~9.

\bibitem{huang2018aclnet}
Jonathan~J Huang and Juan Jose~Alvarado Leanos,
\newblock ``Aclnet: efficient end-to-end audio classification cnn,''
\newblock {\em arXiv preprint arXiv:1811.06669}, 2018.

\bibitem{liu2021swin}
Ze~Liu and et~al.,
\newblock ``Swin transformer: Hierarchical vision transformer using shifted
  windows,''
\newblock in {\em Proceedings of the IEEE/CVF International Conference on
  Computer Vision}, 2021, pp. 10012--10022.

\bibitem{yang2021causalvae}
Mengyue Yang, Furui Liu, Zhitang Chen, Xinwei Shen, Jianye Hao, and Jun Wang,
\newblock ``Causalvae: Disentangled representation learning via neural
  structural causal models,''
\newblock in {\em Proceedings of the IEEE/CVF Conference on Computer Vision and
  Pattern Recognition}, 2021, pp. 9593--9602.

\bibitem{nangi2021counterfactuals}
Sharmila~Reddy Nangi, Niyati Chhaya, Sopan Khosla, Nikhil Kaushik, and Harshit
  Nyati,
\newblock ``Counterfactuals to control latent disentangled text representations
  for style transfer,''
\newblock in {\em Proceedings of the 59th Annual Meeting of the Association for
  Computational Linguistics and the 11th International Joint Conference on
  Natural Language Processing (Volume 2: Short Papers)}, 2021, pp. 40--48.

\bibitem{zhu2018learning}
Boqing Zhu, Changjian Wang, Feng Liu, Jin Lei, Zhen Huang, Yuxing Peng, and Fei
  Li,
\newblock ``Learning environmental sounds with multi-scale convolutional neural
  network,''
\newblock in {\em 2018 International Joint Conference on Neural Networks
  (IJCNN)}. IEEE, 2018, pp. 1--8.

\bibitem{grais2018multi}
Emad~M Grais, Hagen Wierstorf, Dominic Ward, and Mark~D Plumbley,
\newblock ``Multi-resolution fully convolutional neural networks for monaural
  audio source separation,''
\newblock in {\em International Conference on Latent Variable Analysis and
  Signal Separation}. Springer, 2018, pp. 340--350.

\bibitem{zhang2021vidtr}
Yanyi Zhang and et~al.,
\newblock ``Vidtr: Video transformer without convolutions,''
\newblock in {\em Proceedings of the IEEE/CVF International Conference on
  Computer Vision}, 2021, pp. 13577--13587.

\bibitem{gazneli2022end}
Avi Gazneli, Gadi Zimerman, Tal Ridnik, Gilad Sharir, and Asaf Noy,
\newblock ``End-to-end audio strikes back: Boosting augmentations towards an
  efficient audio classification network,''
\newblock {\em arXiv preprint arXiv:2204.11479}, 2022.

\bibitem{izmailov2018averaging}
Pavel Izmailov, Dmitrii Podoprikhin, Timur Garipov, Dmitry Vetrov, and
  Andrew~Gordon Wilson,
\newblock ``Averaging weights leads to wider optima and better
  generalization,''
\newblock {\em arXiv preprint arXiv:1803.05407}, 2018.

\bibitem{tokozume2017learning}
Yuji Tokozume, Yoshitaka Ushiku, and Tatsuya Harada,
\newblock ``Learning from between-class examples for deep sound recognition,''
\newblock {\em arXiv preprint arXiv:1711.10282}, 2017.

\bibitem{elizalde2022clap}
Benjamin Elizalde, Soham Deshmukh, Mahmoud~Al Ismail, and Huaming Wang,
\newblock ``Clap: Learning audio concepts from natural language supervision,''
\newblock {\em arXiv preprint arXiv:2206.04769}, 2022.

\bibitem{zhao2021connecting}
Yanpeng Zhao, Jack Hessel, Youngjae Yu, Ximing Lu, Rowan Zellers, and Yejin
  Choi,
\newblock ``Connecting the dots between audio and text without parallel data
  through visual knowledge transfer,''
\newblock {\em arXiv preprint arXiv:2112.08995}, 2021.

\end{thebibliography}
